# Combining laser ablation and Sol-Gel techniques for the synthesis of nanostructured organic-inorganic matrices

E. Haro-Poniatowski[1,5], C. A. Guarín[1], L. G. Mendoza-Luna[1], L. Escobar-Alarcón[2], J. L. Hernández-Pozos[1], L. I. Vera-Robles[3], P. Castillo[4], F. Cabello[5], J. Toudert[5], F. Chacón-Sánchez[5], M. García-Pardo[5], R. Serna[5], J. Gonzalo[5], J. Solís[5]

*1- Departamento de Física, Universidad Autónoma Metropolitana Iztapalapa, Apdo. Postal 55-534, Cd. de México, México.*
*2- Departamento de Física, Instituto Nacional de Investigaciones Nucleares, Carretera México-Toluca S/N, la Marquesa, Ocoyoacac, Estado de México. C.P.52750 México.*
*3- Departamento de Química, Universidad Autónoma Metropolitana Iztapalapa, Apdo. Postal 55-534, Cd. de México, México.*
*4- Laboratorio Microscopía Electrónica, Universidad Autónoma Metropolitana Iztapalapa, Apdo. Postal 55-534, Cd. de México, México.*
*5- Laser Processing Group, Instituto de Óptica, IO, CSIC, Serrano, 121*
*28006 Madrid (Spain)*

**Abstract**
In this work we report a new and simple method that combines the pulsed laser ablation in liquids (PLAL) and the Sol-Gel techniques to obtain nanocomposite glasses and gelatins. Gold nanoparticles (Au-NPs) are generated by PLAL using the corresponding target. The target is submerged in a transparent liquid solution made previously with tetraetylorthosilicate (TEOS) adding diluted hydrochloric acid as catalyzer. In the case of gelatins commercial gelatin and tap water are used. The laser source is a Nd:YAG laser emitting at 1064 nm, with an energy of 100 mJ and 8 ns pulse duration at 10 Hz repetition rate focused on the target in a 2 mm diameter laser spot. The ablation time is 10 min for the glasses and gelatins. The Au-NPs are uniformly dispersed in the solution. After the ablation process the gels are sealed and stored at room temperature for several days. The samples are characterized by UV-Vis spectroscopy, HRTEM, ellipsometry and AFM microscopy, these measurements reveal optical transparency and a refractive index near 1.45 for the pure glass, whereas a colorful aspect, a refractive index of 1.42, and a small surface roughness of 1.92 nm for the glass containing Au-NPs. In the case of gelatins self-sustained flexible films are obtained.

**Introduction**

Nanocomposite glasses have been around us since ancient times. Nowadays we know that the change in color upon transmission and reflection of light of the exceptional IV[th] century Lycurgus cup fabricated by the romans is due to the fact that it is made of a glass with Au and Ag nanoparticles (NPs) embedded in it [1], forming a nanocomposite. Generally speaking, a nanocomposite is a kind of material in which one phase with nano-sized dimensions is embedded in a ceramic, glass or polymer matrix. A great effort has been made recently to incorporate NPs into a wide variety of matrices that can be liquid or solids, crystalline or amorphous. This incorporation results in changes in the physical and chemical properties of the matrices being in some cases better than the properties of their individual components. Some examples are the improvement of the tribological properties of metals and oils [2, 3]. Nanoparticles of CdTe of different sizes have been dispersed in liquid changing its color progressively from yellow to blue according to their corresponding size [4]. Au nanoparticles dispersed in polymers have been also produced improving their tribological properties [5]. Other applications include the rising field of quantum computing, with great efforts pursued in fabricating quantum dots embedded in different matrices and structures [6]. Additionally, multiple applications in the field of plasmonics require the synthesis of nanostructures in various substrates or matrices [7]. Very recently the idea of embedding nanoparticles into a glassy matrix has been extended to include gelatins as a matrix in order to fabricate eco-friendly optical components such as flexible filters [8].

For a few years we have been involved in the synthesis and development of nanostructured glasses. In particular, our interest has been focused on low melting point metal nanoparticles immersed in high melting point solid matrices. Such systems are of interest because of their potential applications as opto-thermal devices [9, 10]. They display great potential to be used as thermo-optical switches responding to changes in temperature. However, only a few examples can be found in the literature regarding these nanocomposites. To our knowledge only Se, Bi, Pb nanocrystals and more recently CuCl and CuBr have been synthesized successfully in bulk amorphous matrices [11, 12] and references there in.

Finally, it is important to mention that the process of making such glasses is complex, difficult and, furthermore, it is time and energy consuming due to the high temperatures needed in the fabrication process. Consequently, synthesizing such systems "on demand" is not an easy and straightforward process; for example, situations like the fabrication of a glass with elemental Pb NPs inside. Alternative routes have been and are currently under investigation. Pulsed laser deposition (PLD) is a suitable method to fabricate such nanocomposites [13, 14, 15].

An alternative possibility for the fabrication of the desired colloidal systems is the pulsed laser ablation in liquids (PLAL) technique [16]. Nanostructures of different sizes and morphologies including metals, alloys, and oxides have been produced using this technique [17, 18]. An important advantage of this method is its versatility enabling its combination with other fabrication methods in an easy way. In such direction we have synthesized carbon nano-sheets using PLAL under an ultrasound field [19]. However, the reports of PLAL-made nanocomposites are scarce. As mentioned before, nanoparticles have been dispersed by PLAL in polymers and in resins. For example, nail enamels of different colors have been produced by dispersing Au, Ag and Pt nanoparticles by PLD in a nail varnish [20].

On the other hand, for the synthesis of nanocomposite glasses, an alternative and convenient solution is the sol-gel technique. This method has the advantage to be fast and simple and furthermore, the materials are fabricated at room temperature, and the eventual thermal treatments are of the order of only a few hundred °C. The main drawback is that it is not easy to obtain large monoliths. Furthermore, the drying process is long and difficult to optimize. Using this technique nanostructured glasses have been obtained by mixing organo-metallic products and colloids with nanoparticles. However, the variety of organo-metallic products has some limitations such as availability and price. Furthermore, an important advantage of PLAL is its independence from chemical precursors, avoiding the use of substances that can be toxic, or other by-products that can be present in the obtained material.

In the present work we propose to use the combination of the sol-gel and laser ablation techniques to synthesize nanostructured glasses and gelatins [21]. The main requirement is to use a transparent solution at the wavelength used for the ablation. The variety of glassy materials and gelatins that can be fabricated is very broad, while possible limitations could be the availability of the solid target, or the nature of the solvent. In the first approach of this combination of techniques, we have obtained Au nanoparticles embedded in a silica matrix and in a commercial gelatin. The presence of the NPs has been confirmed by different characterization techniques: UV-Vis spectroscopy and TEM microscopy. The synthesis of the nanocomposite glasses reported in the present work is fast and simple compared to other techniques and has several potential applications since the materials can be prepared in bulk and thin film forms. However, a lot of work is needed to optimize several parameters that are determinant in the obtention of the final product. In the case of gelatins, we present some preliminary results in supplementary information [S1]. The fabrication process is simpler than the sol-gel glasses and the results are very similar to those reported in reference [8] except for the method of the nanoparticles incorporation in the gelatine which in our case is by PLAL.

**Experimental set-up**

The experimental procedure is shown schematically in figure 1. The pure (reference) glasses were prepared adding dilute hydrochloric acid, as catalyzer, to the tetraetylorthosilicate (TEOS) with a molar ratio of 1:4 (final concentration of 1%). The catalyzer was prepared by adding 0.18 mL of hydrochloric acid to 1.5 mL of deionized water. This was added gradually to 4.74 mL of TEOS under constant stirring until a homogeneous solution was obtained. The sol-gel solution was prepared 1 hour approximately before the ablation process. After the ablation the samples were transferred into plastic cuvettes of 1.5 mL for UV spectroscopy and covered partially with parafilm to let the evaporation of the solvent occur.

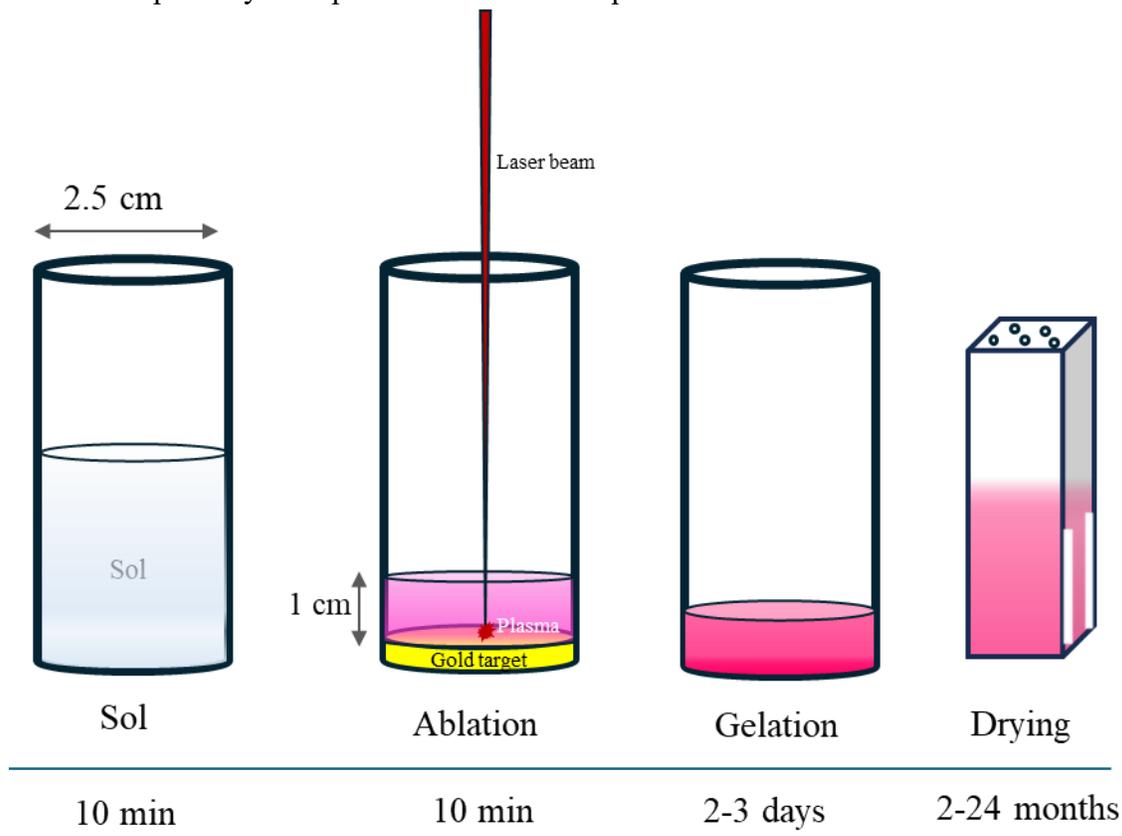

**Figure 1.** Schematic representation of the synthesis of the nanostructured Sol-Gel glasses.

The solution jellifies after 20 to 30 hours and forms a gel which fills the vessel where it is contained. At this point the material starts a shrinking process which can be as important as a 30% loss in weight. This shrinkage occurs in approximately 2 months. After this process the sample ends in a solid block 6 × 2× 10 mm in size.

For the case of gelatins, the procedure is almost identical, however the preparation of the solution is simpler and faster since it involves only gelatin and hot tap water. The ablation process being the same, the gelation time is one hour and drying occurs in 3 days typically.

**Ablation in sol-gel**

The nanocomposite glass was prepared laser-ablating high purity (99.99%) commercial Au target having a diameter of 2.5 cm and thickness of 0.6 cm, submerged in the sol-gel precursor solution. For this purpose, the target was placed inside a glass cylindrical container with an inner diameter slightly larger than the target diameter to fit the target at the bottom. Afterwards, 5 mL of the sol were added forming a layer with a thickness close to 1 cm over the target surface. The laser beam was directed by means of mirrors and focused onto the target surface passing through a lens of 12 cm of focal length. Laser ablation was carried out using the fundamental emission of a Nd:YAG laser at a wavelength of 1064 nm and 8 ns of pulse duration, working at a repetition rate of 10 Hz. Experiments were performed at a laser energy of 100 mJ in a 2mm diameter laser spot. The ablation time was typically 10 min, this value was determined empirically when observing that the sol turned from transparent to a pink color. The longer the ablation time the higher is the density of nanoparticles, however the ablation efficiency decreases with time because the suspended Au nanoparticles have a screening effect on the incident radiation. In the case of gelatins, 1.6 g of commercial pure gelatin were mixed with 62.5 ml of boiling water from tap under constant stirring for 20 min. Once the temperature decreased to 40°C the mixture was poured into a vessel for the laser ablation process. The same configuration is used for TEOS and gelatins, however, the preparation parameters are different since the ablation process is more efficient in the gelatin solution than in the TEOS solution therefore for the same energy density the ablation time is reduced significantly typically 50%. In both cases the presence of gold nanoparticles can be observed readily since the solution exhibits a pink color as shown in figure 2 and figure S.1.1 in supplementary information S1.

**Results and discussion**

First of all, the optimum experimental conditions have to be determined. The main parameters involved in the synthesis of nanocomposite glasses combining the PLAL and Sol-Gel techniques are the following: a) Transparency of the sol, b) Viscosity, c) Ablation Time (viscosity increases with time), d) Liquid amount over the target, e) Energy density, f) Wavelength, g) Post-preparation thermal treatments. Furthermore, the optimization of these parameters in some cases can involve the change of the original formula used for the preparation of the sol as discussed later. Most of these parameters have been identified in previous works [20, 21]. A very sensible one is viscosity, since it is directly related to the proper dispersion of the nanoparticles this aspect needs to be thoroughly investigated in future experiments.

**Sol-Gel glasses**

It is important to start the ablation within one hour after the preparation of the sol. During ablation upon increasing viscosity the nanoparticles tend to accumulate vertically in a columnar way at the laser spot perpendicularly to the target. The effect of increasing viscosity with time is easily observable by eye, however a suitable experimental procedure is needed for obtaining quantitative information. This investigation is out of scope of the present work. The liquid amount is also crucial for the laser ablation process, a thin layer of liquid results in important splashing. In the present experiments we have used a liquid layer typically 1 cm above the surface of the target. As soon as the ablation process takes place the production of gold nanoparticles can be detected visually. In figure 2, for coloration comparison, the obtained gel is presented (figure 2a) together with a colloid of Au-NPs obtained by PLAL in distilled water (figure 2b) in the same experimental conditions.

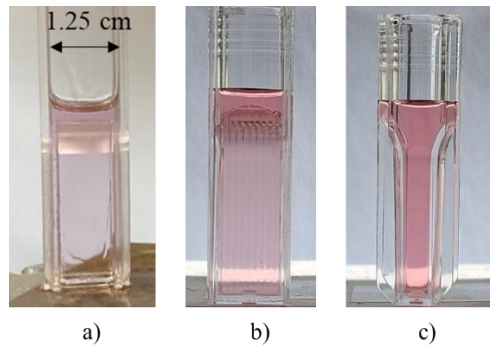

**Figure 2.** a) as prepared Au- NPs in the sol-gel solution, b) Au- NPs in distilled water, c) lateral view of b).

These Gels were immediately characterized by UV-Vis spectroscopy with a Varian Cary 2000 double beam spectrophotometer in the transmission configuration at normal incidence. The absorbance spectra were obtained from the measured transmittance spectra following Absorbance = 1 – Transmittance, a transmittance of 1 meaning that 100% of the light intensity passes through the medium. The same setup, configuration, and measurement method were used for all the UV-vis spectroscopy measurements of this paper.

Before ablation a transparent reference sample is withdrawn from the sol. UV-Vis absorbance spectra of the reference gel, gel with Au-NPs as prepared and gel with Au-NPs after 10 days of drying are presented in figure 3. The blue spectrum corresponds to the reference sample and as expected no absorbance features are observed in the visible range. The spectrum depicted in red shows a band around 540 nm that reveals the presence of Au nanoparticles in the gel and is attributed to the surface plasmon resonance of Au. The 540 nm resonance wavelength corresponds to NPs with a non-retarded dipolar resonance. If we assume - as a first approximation – that the NPs are monodisperse with a spherical shape, and that the refractive index of the surrounding medium is 1.44, Mie calculations (Supporting Information S2) suggest that the 540 nm peak position corresponds to a NPs diameter between 40 and 50 nm. Note also that the measured plasmon band spans over the 520 to 800 nm shape and displays an asymmetric shape with a broader extension toward longer wavelengths. This contrasts with Mie simulations for a given NP size and suggests that the NPs have a broad polydisperse size and/or shape distribution, with bigger and/or non-spherical NPs resonating on the long wavelength side of the spectrum [22]. Finally, from effective medium simulations (Maxwell-Garnett model) of the measured absorbance spectrum, we estimate the volume fraction of Au NPs to be very small, in the $10^{-4}$ % to $10^{-3}$ % range.

The black spectrum was recorded after 10 days of preparation. The striking feature in this spectrum is that the band corresponding to the well-defined Au plasmon resonance has disappeared and an intense band is peaking at 324 nm. Here we provide a preliminary discussion about why this well-defined plasmon has disappeared. While no well-defined plasmon can be seen in the spectrum, there is a broad background, which might result from the presence of bigger NPs – with sizes of hundreds of nm – that do not show well-defined plasmons as supported by the Mie simulations shown in the supplementary information [S2]. In other words, the NPs might have formed bigger structures within the matrix, either by coalescence or Ostwald ripening. Another, yet less likely possibility, is that a significant amount of Au has migrated toward the surface where it formed big NPs, as has been reported in other studies for Ag in glasses and thin films [23, 24]. One might also think that the Au NPs have chemically reacted with the matrix (e.g. Au-$OC_2H_5$ complexes) or lost their plasmonic character due to intense chemical interface damping [25]. However, such effects usually have a strong impact only in the case of few-nm NPs, much smaller than in the present case. Note also that the appearance of the band at 324 nm is somewhat simultaneous to the disappearance of the Au plasmon band. This appearance is likely due to changes in the matrix, as Au NPs cannot display plasmon resonances in the UV. Further characterization, beyond the scope of this paper, is needed to definitely address the origin of the observed trends.

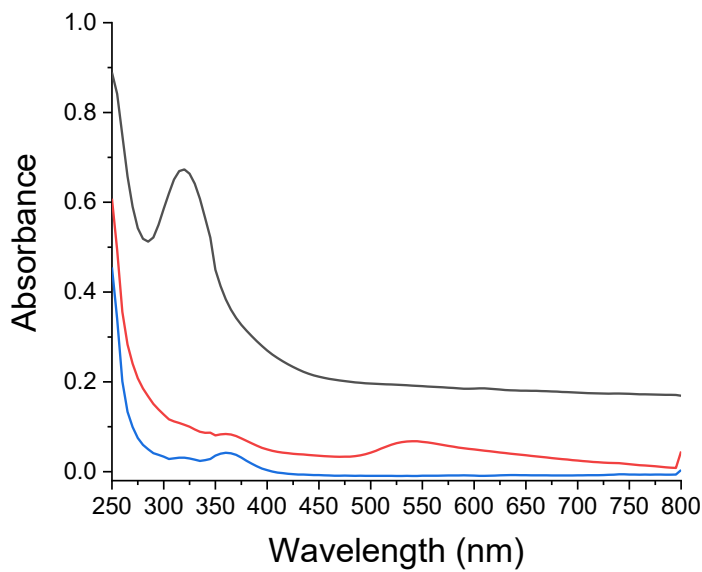

**Figure 3.** Absorbance spectra of the gel. Blue curve corresponds to the reference (without Au) sample. Red curve: sol-gel obtained after the ablation of the Au target showing the presence of the plasmon band at 540 nm. Black curve corresponds to the same sample with Au-NPs after 10 days, a strong peak at 324 nm is clearly observable and the plasmon band has disappeared.

**Drying and shrinking processes**

The drying procedure appears to be crucial for the final result and is still under investigation. There are many publications regarding this aspect of the synthesis of sol-gel glasses. [27, 28]. However, at this point this issue is out of the scope of the present work. Once the sol has been poured into the UV cuvette the drying process starts, we have measured the weight loss as a function of time as shown in figure 4.

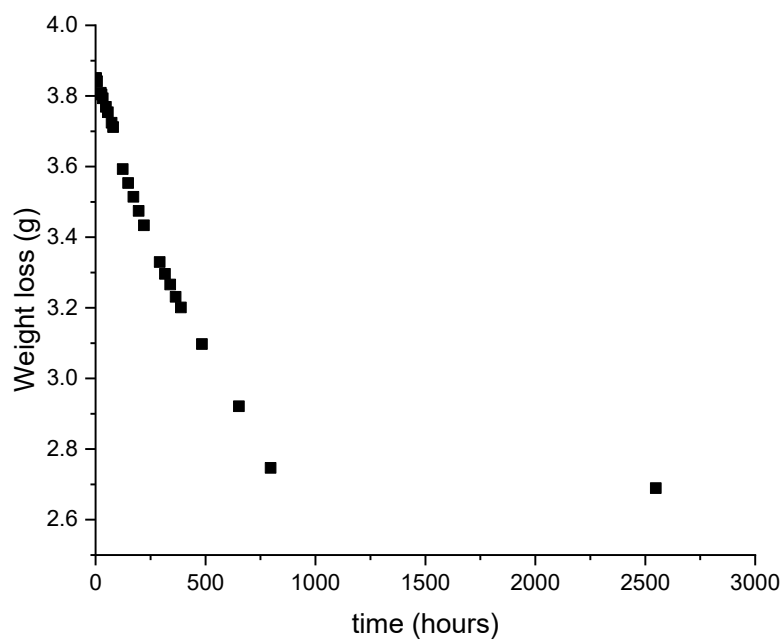

Figure 4 Weight-loss of a sol-gel sample in a cuvette as a function of time. After approximately 3 months of drying the sample has shrunk and lost about 30% of its original weight.

In the first hour, the evaporation occurs quite fast and if the vessel is not partially sealed the probability of fracture once the material solidifies is very high. We have tried to optimize the drying process by changing the vessels which determine the morphology and size of the glasses and furthermore the sealing procedure of the vessels during the drying procedure. The vessels are covered with a plastic film with some holes to allow evaporation. Best results have been obtained using UV-Vis spectroscopy cuvettes designed for 1.5 mL sealed with parafilm with some small holes in it. Just after preparation the sol-gel solution occupies most of the sample volume of the cuvette (photograph not shown here). In figure 5 we show the drying-shrinking process. Figure 5(a) and 5(b) shows the sample after 1 month and after 2 months of preparation respectively. The final product is shown in 5(c) after 3 months of preparation and after some slight polishing of the top. This is a quite hard transparent block of glass 6 ×2×10 mm in size. At this stage the obtained sample is very robust and can be easily manipulated but it is certainly not completely dry. We have examined samples which have dried for as long as two years and although the optical properties of the sample exhibit small changes we cannot assure that they are completely dry.

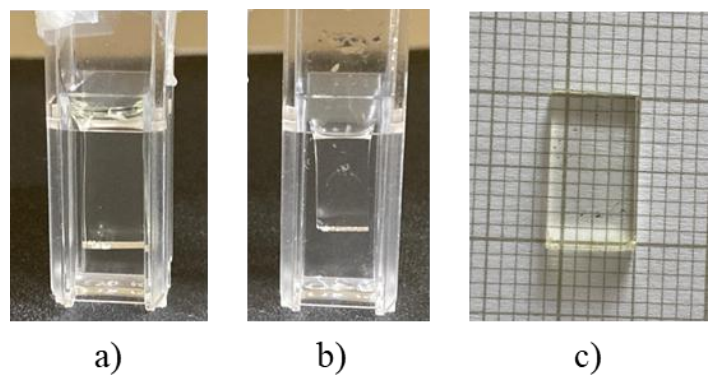

a) b) c)

**Figure 5** Shrinking process after: (a) 1 month and (b) 2 months; (c) block obtained after 3 months of drying (images not in scale).

Regarding the final block shown in figure 5 (c) an AFM image reveals a very flat surface as shown later. The only surface that must be polished is the superior one which maintains the block vertically in the cuvette. In the case of gelatins, the drying and shrinking processes are described in in supplementary information S1.

**Thermal treatments**

In many cases and depending on the preparation variables used for the synthesis of the nanocomposites, once they have almost completely dry in the form of blocks or thin flat pieces, these can be thermally treated to retrieve the original pink color and the characteristic Au plasmon resonance near 530 nm. Thermal treatments at heating rates ranging from 1 °C/min to 10 °C/min were used. Best results are obtained for slow heating rates and depend on the shape and thickness of the samples. The results and UV-Vis spectra are shown in figure 6. Figure 6(a) shows a photograph of a piece of glass before (left) and after (right) a thermal treatment at 377°C for 5 hours at a heating rate of 3°C/min. Figure 6 (b) shows the absorbance spectra of a nanocomposite glass with Au-NPs before and after two thermal treatments at 377°C and 500°C for 5 hours at a heating rate of 3°C/min, the Au plasmon at 530 nm is clearly visible over a spectrally broad background; it is included the spectrum of a reference glass after thermal treatment at 500°C. This suggests that, while big, few-hundred nm Au-NPs dominate the broadband optical response of the annealed glass, the annealing has also led to the formation of small, few-nm Au-NPs, perhaps from nucleation-growth involving atoms that remained isolated in the matrix prior to annealing.

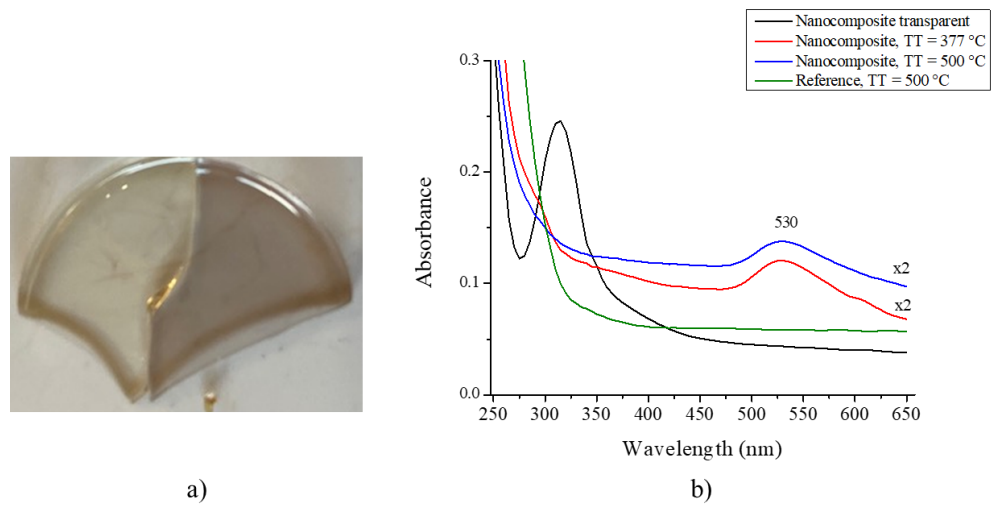

a) b)

**Figure 6** (a) shows a photograph of a piece of glass before (left) and after (right) a thermal treatment at 377°C. Figure 6 (b) shows the absorbance spectra in the 250-650 nm range for the reference glass and after two thermal treatments at 377°C and 500°C.

However, it appears that after a very long drying process the original plasmon resonance peak remains. In figure 7 we show photographs of a reference sample (figure 7a) and a nanocomposite with gold nanoparticles (figure 7b) that have been drying for 6 months. The sample on the left side is illuminated in reflection whereas the sample on the right side is illuminated in transmission. For the sample containing Au-NPs the change in color is evident. Figures 7c and 7d show a sample after 2 years of drying in transmission and reflection respectively resembling the well-known dichroic effect observed in the Lycurgus cup. In the supplementary information [S3] we show a video to illustrate this change of color in real time for the case of the sample that has been drying for two years. As mentioned before, a thermal treatment helps to accelerate the nanocomposite color recovery which is very slow if the glass is left drying for a long time.

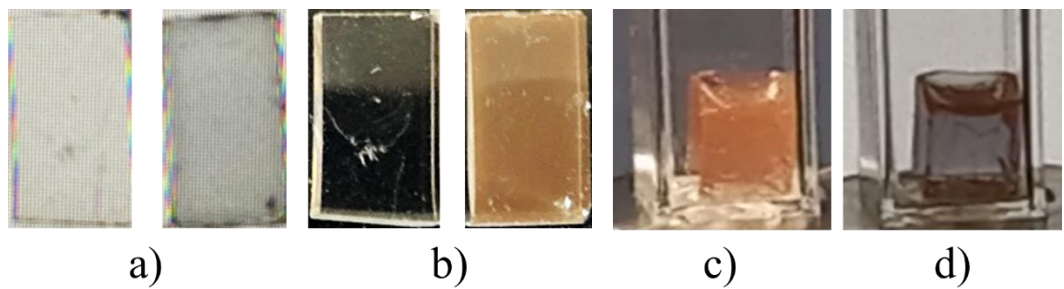

a) b) c) d)

Figure 7. Photographs in transmission and reflection of a reference (a) and a sample with Au-NPs (b) that have been drying for 6 months; and a nanocomposite with Au-NPs after two years of drying in transmission (c) and reflection (d).

**HRTEM Characterization**

To confirm the presence of Au-NPs in the obtained nanocomposite glasses we have obtained HRTEM images. For this purpose, the sample was prepared by mechanical grinding with an agate mortar. The powder was placed in a vial with alcohol and kept in ultrasound for 30 min to disperse and homogenize it. Afterwards a drop of the obtained suspension was deposited on a formvar coated Cu grid allowing the solvent to evaporate completely at room temperature. Figure 8 shows a high-resolution image in which nanoparticles with an interplanar distance of 0.233 nm corresponding to the (111) family planes, characteristic of gold, are observed. EDS spectra confirms that these are Au nanoparticles. A sample was analyzed by high angle annular dark field (HAADF) technique in a JEOL HRTEM microscope and from some images (not shown here) sizes ranging from 10 to 43 nm in diameter were measured with an average of 20 nm, determined analyzing 22 Au-NPs.

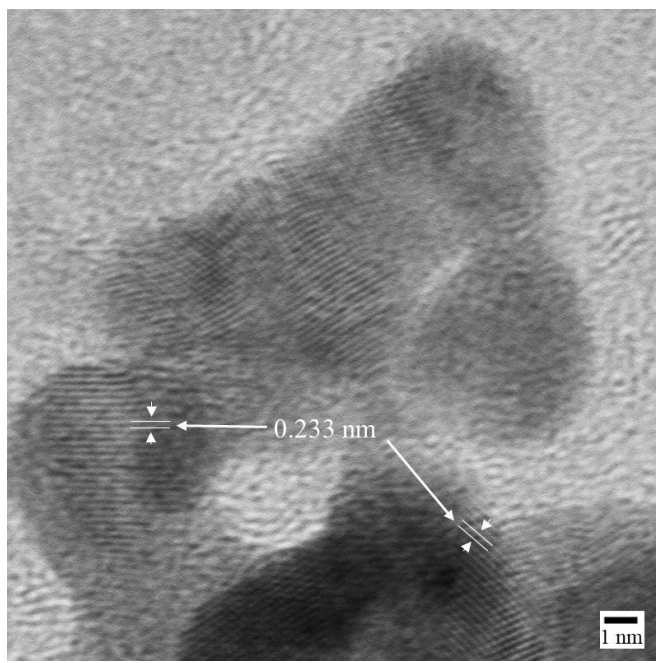

Figure 8. HRTEM image of a nanocomposite sample after two years of drying.

**AFM Characterization**

To characterize the surface topography of the synthesized glasses, AFM measurements were performed via non-contact mode measurements. On figure 9 we can appreciate the smoothness of the surface, with most height values in the ± 4nm range and a mean square roughness of 1.92 nm. This confirms that the glasses have a smooth surface that will not play any role in the optical characterization, as well as opening a pathway for applications as a functional substrate. Interestingly, we can also observe some cracks in the surface that could be associated with the shrinking process described earlier.

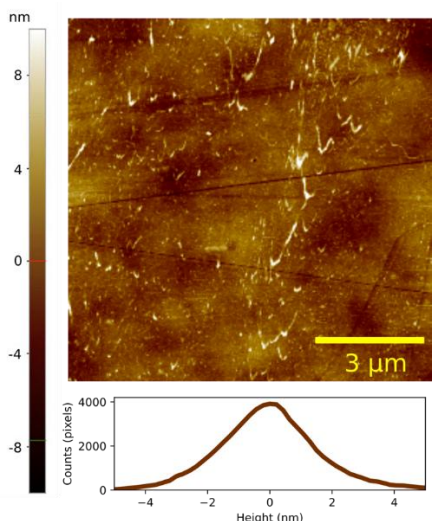

Figure 9. AFM image of a pure glass sample after 6 months of drying.

Optical Characterization:

The optical characterization of the glass without Au-NPs was performed in a spectroscopic ellipsometer, covering from the UV to the near-IR wavelength range (300–1700 nm). The results for n and k are presented in figure 10. It shows a refractive index close to 1.45 in the UV-Vis-NIR, while k remains below the limit of detection of the ellipsometer (0.01). Measurements on glasses containing Au-NPs show similar n and k values, because the volume fraction of NPs in the nanocomposite glass being too small to sizeable modify its effective optical constants compared with the glass without NPs.

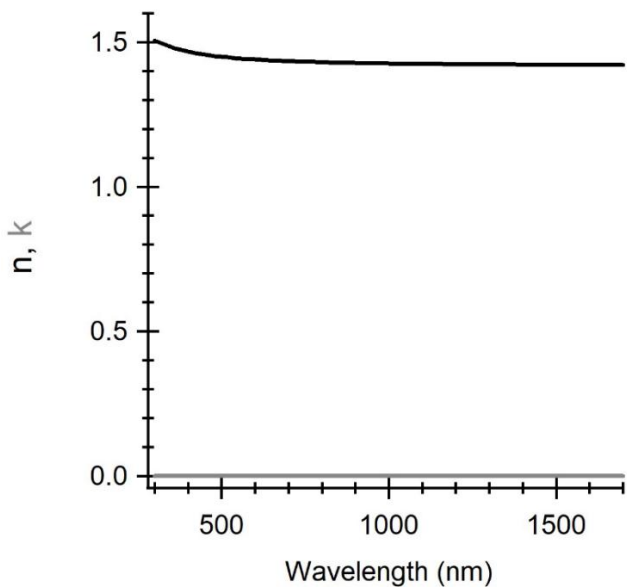

Figure 10. Ellipsometry results for the n and k values in the range from 300 to 1700 nm of a glass without Au-NPs.

**Conclusions**

In this investigation we have shown that it is possible to synthesize nanocomposite glasses and gelatins by combining the Sol-Gel and the pulsed laser ablation in liquids techniques. Using this combination we have been able to embed Au nanoparticles in a silicate matrix as well as in commercial gelatins. The synthesis process and the obtained nanocomposites strongly depend on the laser ablation parameters and on the drying procedure. The experimental approach we report can be implemented using a broad range of targets and precursors to achieve a wide family of nanocomposite materials with tailored optical and other properties with potential applications in photonics.

**Acknowledgments**

We are grateful to the Universidad Autónoma Metropolitana and SECIHTI through grant INFR-2015-255579 ''Infraestructura para Laboratorio de Interacción Radiación Materia y Espectroscopía de Procesos Ultrarrápidos'' for their funding. C. A. G. and L. G. M. L. are grateful to SECIHTI for funding through grant 683 in the framework of the ''Investigadores por México'' scheme. E. H. P. thanks the Secretaría de Ciencia, Humanidades, Tecnología e Innovación (SECIHTI) of México for financial support. The ININ support through the project CB-004 is acknowledged. Also, we acknowledge the help in the early stages of this investigation the help of J. G. Morales-Méndez, T. Aguilar-Sánchez, A. Limas-Escobar and A. Hernández-Gordillo. We thank the Laboratorio Microscopía Electrónica, Universidad Autónoma Metropolitana Iztapalapa for the TEM and HRTEM studies.

**Conflict of interest**

The authors declare that they have no conflict of interest.

**Contribution statement**

EHP: Writing – original draft, Supervision, Methodology, Investigation, Formal analysis, Conceptualization. CG: Investigation, Methodology. LGML: Investigation, Methodology. LEA: Methodology, Investigation, Formal analysis. JLHP: Investigation, LIVR: Methodology, Investigation, PC: Investigation, Formal analysis. FC: Investigation, JT: Investigation, Formal analysis, Methodology, Investigation. FCS: Investigation, Formal analysis. MGP: Investigation. RS Methodology, Investigation. JG: Methodology. JS: Methodology.

**Data availability statement**

The datasets generated during and/or analysed during the current study are available from the corresponding author on reasonable request.

# Supplementary information:

**S1 Gelatins results**

As in the case of the nanostructured glasses, the PLAL procedure has to start before the gelatine has jellified. As said before just after preparation the gelatin is liquid and close to 70°C in temperature. Jellification occurs around 18°C therefore we have incorporated the Au NPs into the gelatin by PLAL at a temperature around 40 °C and jellification occurred after a few hours.

**Drying and shrinking processes in gelatins.**

The drying and shrinking are different in the case of gelatins compared to the sol-gel glasses. The most interesting results were obtained for gelatins that were dried in petri dishes. The nanostructured hot liquid was poured into the petri dish and had a height of 5 mm. After 10 days drying a flexible thin film 20-50 microns thick is obtained.

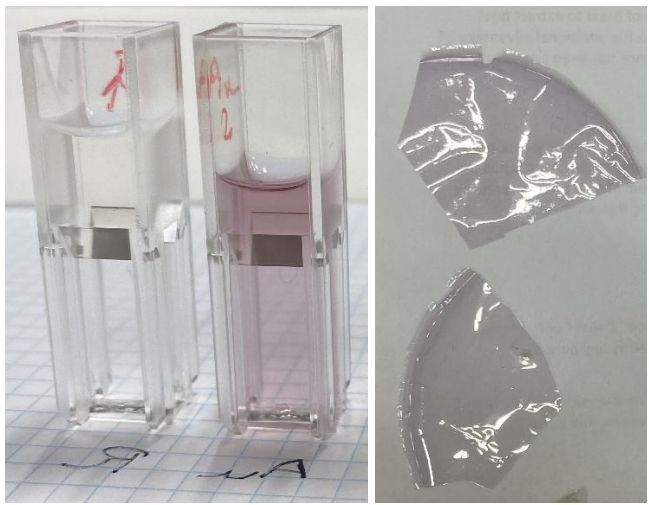

**Figure S1.1** Left: as prepared gelatin and as prepared gelatin after PLAL (with Au NP-s). Right: Dried gelatin films with Au NP-s.

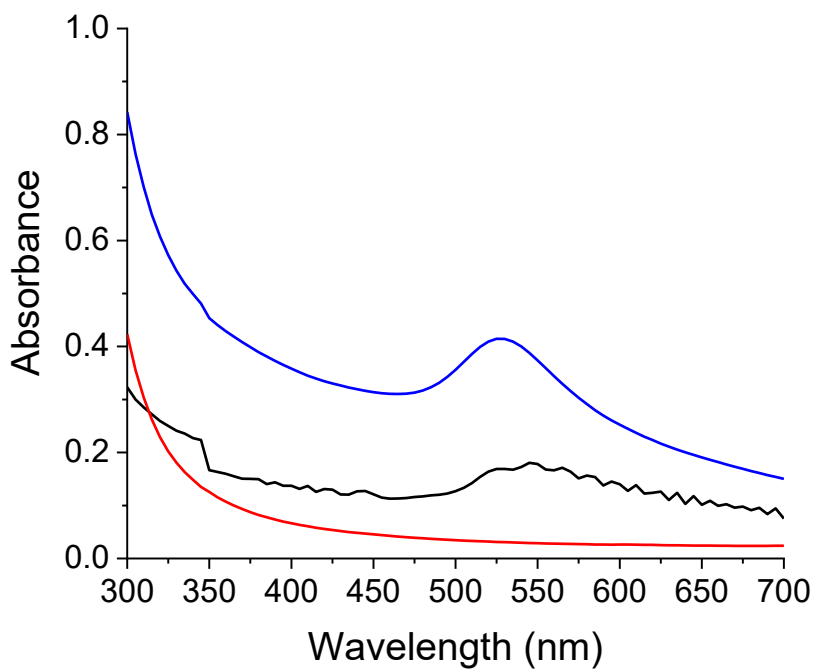

**Figure S.1.2** Absorbance spectra of the gel (Absorbance = 1 – Transmittance, a transmittance of 1 meaning that 100% of the light intensity passes through the medium). Blue curve corresponds to the as prepared gelatin sample with Au-NP's sample. The black curve corresponds to the gelatin dried sample (thin film of gelatin with Au NP's. Some interference effects are clearly observable. Red curve: as prepared reference gelatin.

## S2. Mie simulations

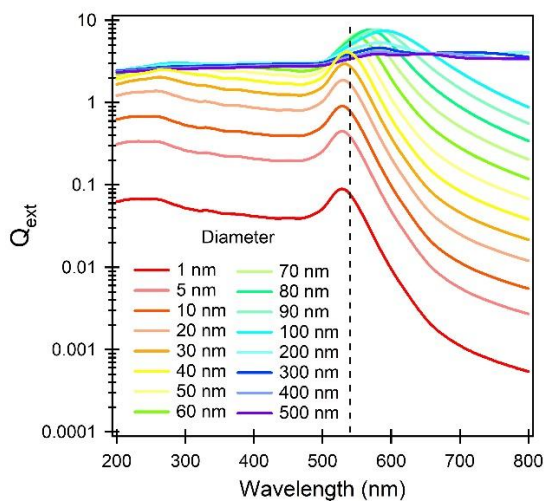
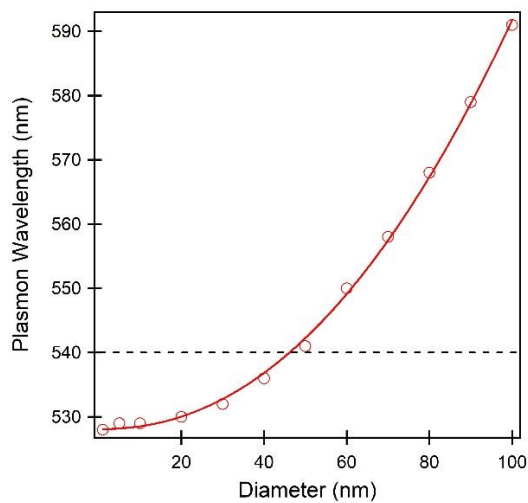

**Figure S2.1** Mie simulations for a spherical Au-NPs in a matrix. The NP diameter was varied between 1 and 500 nm. The refractive index of the matrix was 1.44. The obtained extinction spectra ($Q_{ext}$ vs wavelength) are plotted in a log scale on the left graph. The vertical dashed line represents the 540 nm wavelength. The plasmon resonance wavelength is plotted vs the NP diameter on the right graph, where the horizontal dashed line represents the 540 nm wavelength.

**S3 Video**

**S3.1** Video that illustrates the change of color upon illumination in reflection and transmission in real time for the case of the glass sample that has been drying for two years.

Video available on: https://youtu.be/WKpvlA15s68